\documentclass{pasj00}
\newcommand{\bm}[1]{\mbox{\boldmath $#1$}}
\begin{document}
\SetRunningHead{Hashizume et al.}
{Wide-angle Outflows from Super-critical Accretion Disks}

\title{Radiation hydrodynamics simulations of wide-angle outflows from super-critical accretion disks around black holes}

\author{Katsuya \textsc{Hashizume} %
  \thanks{Example: Present Address is $katsuya.hashizume@gmail.com$}}
\affil{School of Physical Sciences,Graduate University of
Advanced Study (SOKENDAI), Shonan Village, Hayama, Kanagawa 240-0193, Japan}
\email{katsuya.hashizume@nao.ac.jp}

\author{Ken \textsc{Ohsuga}}
\affil{
National Astronomical Observatory of Japan, Osawa, Mitaka, Tokyo
181-8588, Japan
\\
School of Physical Sciences,Graduate University of
Advanced Study (SOKENDAI), Shonan Village, Hayama, Kanagawa 240-0193,
Japan}
\email{ken.ohsugab@nao.ac.jp}

\author{Tomohisa {\sc Kawashima}}
\affil{Key Laboratory for Research in Galaxies and Cosmology, 
Shanghai Astronomical Observatory, Chinese Academy of Science, 
80 Nandan Road, Shanghai 200030, China
\\
National Astronomical Observatory of Japan, Osawa, Mitaka, Tokyo
181-8588, Japan}
\email{kawashima-t@shao.ac.cn}

\and
\author{Masaomi {\sc Tanaka}}
\affil{
National Astronomical Observatory of Japan, Osawa, Mitaka, Tokyo
181-8588, Japan
\\
School of Physical Sciences,Graduate University of
Advanced Study (SOKENDAI), Shonan Village, Hayama, Kanagawa 240-0193,
Japan}
\email{masaomi.tanaka@nao.ac.jp}

%

\KeyWords{accretion, accretion disks---black hole physics---
  ISM: jets and outflows---X-rays: galaxies} 

\maketitle

\begin{abstract}
By performing two-dimensional radiation hydrodynamics
simulations with large computational domain of 5000 Schwarzschild radius,
we revealed that wide-angle outflow is launched 
via the radiation force from 
the super-critical accretion flows around black holes.
The angular size of the outflow,
of which the radial velocity ($v_r$) is over the escape velocity
($v_{\rm esc}$),
increases with an increase of the distance from the black hole.
As a result, the mass is blown away with speed of $v_r > v_{\rm esc}$
in all direction except for the very vicinity 
of the equatorial plane, $\theta=0^\circ-85^\circ$,
where $\theta$ is the polar angle.
The mass ejected from the outer boundary per unit time by the outflow
is larger than the mass accretion rate onto the black hole,
$\sim 150 L_{\rm Edd}/c^2$,
where $L_{\rm Edd}$ and $c$ are
the Eddington luminosity and the speed of light.
Kinetic power of such wide-angle high-velocity outflow
is comparable to the photon luminosity and 
is a few times larger than the Eddington luminosity.
This corresponds to 
$\sim 10^{39}-10^{40}\ {\rm erg\ s^{-1}}$
for the stellar mass black holes.
Our model consistent with the observations of 
shock excited bubbles observed in some ultra-luminous X-ray sources 
(ULXs),
supporting a hypothesis that 
ULXs are powered by the super-critical accretion 
onto stellar mass black holes.
\end{abstract}

\section{Introduction}
\label{intro}
Astrophysical black holes,
such as active galactic nuclei,
black hole binaries, 
and possibly gamma-ray bursts,
are thought to be 
powered by disk accretion flows (accretion disks).
From a theoretical point of view,
the disks are known to exhibit three distinct modes 
according to the mass accretion rate;
standard disk (\cite{Shakura-Sunyaev1973}1973), 
radiatively inefficient accretion flow
/advection dominated accretion flow 
(\cite{Ichimaru1977}1977; \cite{Narayan-Yi1994}1994),
and slim disk (\cite{Abramowicz1988}1988).
The slim disk appears when the accretion rate is much larger than the
critical rate, $L_{\rm Edd}/c^2$, where $L_{\rm Edd}$ and $c$ are
the Eddington luminosity and the speed of light.
\citet{Sadowski2009} has reported that the
standard disk turns into the slim disk at the accretion rate of
$L_{\rm Edd}/(c^2 \eta_a)$,
where the efficiency, $\eta_a$, is around 0.1 or less
and is a function of a spin parameter of the black hole.
%
The slim disk can shine above the Eddington luminosity
in contrast to that the disk luminosities of 
standard disk and radiatively inefficient accretion flow
are less than the Eddington luminosity.
The slim disk basically succeeded in reproducing the
observed features of ultra-luminous X-ray sources (ULXs) 
and narrow-line Seyfert 1 galaxies,
which are candidates of near- or super-critical flows
(\cite{Watarai2001}2001; \cite{Mineshige2000}2000;
\cite{Kawaguchi2003}2003).

Outflow from the slim disk is of interest to recent 
observational and theoretical studies.
It has been reported by recent observations
that powerful outflows are ejected 
from the central engine of ULXs.
Some of the bubbles/nebulae around ULXs 
are shock-excited via the outflow,
of which the kinetic power is estimated as
$\sim 10^{39-40}\, {\rm erg\ s^{-1}}$
(\cite{Pakull-Mirioni2002}2002; \cite{Pakull-Mirioni2003}2003;
\cite{Grise-Pakull-Motch2006}2006; \cite{Abolmasov2007}2007;
\cite{Cseh2012}2012).
Since the slim disk is basically one-dimensional model,
the launching outflow is not taken into consideration.
Two-dimensional version of the slim disk
is studied by radiation hydrodynamics (RHD) simulations
(\cite{Eggum1988}1988; \cite{Okuda2002}2002; \cite{Ohsuga2006}2006;
\cite{Kawashima2009}2009)
and radiation magnetohydrodynamics simulations
(\cite{Ohsuga2009}2009; \cite{Takeuchi2010}2010; 
\cite{Ohsuga-Mineshige2011}2011; 
\cite{2014MNRAS.439..503S}).
\cite{Ohsuga2005}(2005; hereafter O05)
for the first time demonstrated
the quasi-steady super-critical accretion is feasible.
The luminosity and the mass accretion rate exceed
the Eddington luminosity and the critical rate.

O05 also revealed that 
the outflow is launched from the disk surface
via the radiation force for Thomson scattering.
High-velocity outflow, which is defined as 
the matter being blown away with speed of $>v_{\rm esc}$,
appear around the rotation axis,
where $v_{\rm esc}$ is the escape velocity.
However, at wide off-axis angle,
huge amount of mass 
passes out through the outer boundary at $500r_{\rm S}$, 
with $r_{\rm S}$ being Schwarzschild radius,
with velocity of $<v_{\rm esc}$ (low-velocity outflow).
Behavior of such low-velocity outflow 
is significant issue.
If the outflowing matter continues to be accelerated 
and exceeds the escape velocity at a great distance,
the angular size of the high-velocity outflow 
would increase significantly.
Otherwise the gas comes back to the vicinity of the black hole
and mass accretion rate would go up.
In the former case, 
powerful outflows might excite the interstellar medium via shock,
reproducing the shock-excited bubbles observed around ULXs.
The mass accretion might be interrupted by 
the ram pressure and/or shock heating,
preventing quasi-steady super-critical accretion.
However, global structure and dynamics of the outflows at the large distance
have not been investigated previously.
This is a motivation for the present study.

Here, we perform two-dimensional RHD simulations of
super-critical accretion flows and outflows.
We employ large computational box of
$5000 r_{\rm S}$.
We explain basic equations and the numerical methods in 
\S\ref{basic eq}, 
and display the structure of the 
wide-angle powerful outflows \S\ref{result}.
Finally, \S\ref{discussion} 
and \S\ref{conc} are devoted to discussion and conclusions.

\section{Basic equations and numerical method}
\label{basic eq}
In this section, we describe basic equations of RHD 
and our numerical method.
We employ spherical polar coordinates $(r,\ \theta,\ \varphi)$, 
where $r$ is the radial distance, 
$\theta$ is the polar angle, 
and $\varphi$ is the azimuthal angle.
Our numerical method is same as O05 except for
the size of the computational domain.
While they set the size of the domain to be 500$r_{\rm S}$,
we extended it to 5000$r_{\rm S}$ 
in order to investigate the motion of the outflows 
at the outer regions of $r>500r_{\rm S}$.

Since we assume axial symmetry as well as reflection symmetry
with respect to the equatorial plane,
the computational domain consists of spherical shells of
$r_{\rm in}\ltsim r\ltsim r_{\rm out}$ 
and $0 \ltsim \theta \ltsim \frac{\pi}{2}$,
where $r_{\rm in}$ and $r_{\rm out}$ are set to be
$3r_{\rm S}$ and $5000r_{\rm S}$.
This domain is divided into $N_r\times N_\theta=140\times 96$ grid
cells,
though O05 employs $N_r\times N_\theta=96\times 96$ grid points.
The $N_r$ grids along the radial direction are equally spaced logarithmically,
while the $N_\theta$ grids 
are equally distributed as satisfying $\Delta\cos\theta=1/N_\theta$.
Here, because the number of 
grid points in radial direction within $r=500r_{\rm S}$ is 96,
the resolution of the present work 
is the same as that of O05 near the black hole.

By an explicit-implicit finite difference scheme on the Eulerian grids,
we numerically solve the RHD equations including viscosity terms,
the equation of continuity,
\begin{equation}
\frac{\partial\rho}{\partial t}+\nabla\cdot(\rho {\bm v})=0,
 \label{cont}
\end{equation}
the equation of motion,
\begin{eqnarray}
\frac{\partial(\rho v_r)}{\partial t} 
 &+& \nabla\cdot(\rho v_r{\bm v}) \nonumber\\
 = &-&\frac{\partial p}{\partial r}
 + \rho\left[\frac{v_\theta^2}{r}+\frac{v_\varphi^2}{r}-\frac{GM}{(r-r_{\rm S})^2}\right]
 + \frac{\chi}{c}F_0^r,
 \label{momr}
\end{eqnarray}
\begin{equation}
\frac{\partial(\rho rv_\theta)}{\partial t} + \nabla\cdot(\rho rv_\theta {\bm v})
 = -\frac{\partial p}{\partial\theta} + \rho v_\varphi^2\cot\theta +
\frac{\chi}{c}rF_0^\theta,
\label{momth}
\end{equation}
\begin{eqnarray}
\frac{\partial(\rho r\sin\theta v_\varphi)}{\partial t}
 &+& \nabla\cdot(\rho r\sin\theta v_\varphi{\bm v}) \nonumber\\
 &=& \frac{1}{r^2}\frac{\partial}{\partial r}
 \left[
   \eta r^4 \sin\theta 
   \frac{\partial}{\partial r}\left(\frac{v_\varphi}{r}\right)
 \right],
 \label{momphi}
\end{eqnarray}
the energy equation of the gas,
\begin{eqnarray}
\frac{\partial e}{\partial t}&+&\nabla\cdot(e{\bm v}) \nonumber\\
 &=& -p\nabla\cdot{\bm v} - 4\pi\kappa B
 + c\kappa E_0
 + \eta\left[r\frac{\partial}{\partial r}\left(\frac{v_\varphi}{r}\right)\right]^2,
 \label{gase}
\end{eqnarray}
the energy equation of the radiation,
\begin{eqnarray}
\frac{\partial E_0}{\partial t} &+& \nabla\cdot(E_0{\bm v}) \nonumber\\
 &=& -\nabla\cdot{\bm F}_0 - \nabla{\bm v}:{\bf P}_0
 + 4\pi\kappa B - c\kappa E_0. 
 \label{rade}
\end{eqnarray}
%
Here $\rho$ is the gas mass density, ${\bm v}=(v_r,\ v_\theta,\ v_\varphi)$ is the
velocity, $p$ is the gas pressure, 
$M$ is the black hole mass,
$e$ is the internal energy density of the gas,
$B$ is the blackbody intensity, $E_0$ is the radiation energy density,
${\bm F}_0=(F^r_0,\ F^\theta_0)$ is the radiation flux, ${\bf P}_0$ is
the radiation pressure tensor, $\eta$ is the dynamical viscosity coefficient,
$\kappa$ is the absorption opacity, and $\chi(=\kappa+\rho\sigma_{\rm T}/m_p)$
is the total opacity, where $\sigma_{\rm T}$ is the Thomson scattering cross
section and $m_p$ is the proton mass.
Throughout the present study, we employ $M=10M_\odot$.
For the absorption opacity, $\kappa$,
we consider the free-free absorption 
and the bound-free absorption for solar metallicity
(\cite{Hayashi1962}1962; \cite{Rybicki1979}1979; see also O05).

In addition, we use a equation of state,
\begin{equation}
p=(\gamma-1)e,
\end{equation}
where $\gamma (=5/3)$ is the specific heat ratio.
The temperature of the gas, $T$, can be calculated from
\begin{equation}
p=\frac{\rho k_{\rm B}T}{\mu m_p},
\end{equation}
where $k_{\rm B}$ is the Blotzmann constant and $\mu (=0.5)$ 
is the mean molecular weight.

We apply the flux-limited diffusion (FLD) approximation
(\cite{Levermore-Pomraning1981}1981).
Then, the radiation flux, ${\bm F}_0$,
and the radiation pressure tensor, ${\bf P}_0$,
are expressed in terms of the radiation energy density
(see O05).
%
We consider only $r\varphi$-componet in the viscous stress tensor.
The dynamical viscosity coefficient is given by
$\eta=\alpha(p+\lambda E_0)/\Omega_{\rm K}$,
where $\alpha (\equiv 0.1)$ is the viscosity parameter, 
$\Omega_{\rm K}$ is the Keplerian angular speed, 
and $\lambda$ is the flux limiter,
which becomes $1/3$ in the optically thick limit
and null in the optically thin limit.
Here we note that the momentum conservation is 
not strictly accurate in the present simulations. 
This is because that, in the FLD approximation, 
the radiation flux is simply given as a function of 
the gradient of the radiation energy without 
being solved by conservative form. 
Such problem should be resolved by numerical simulations 
with M-1 closure method 
(\cite{2007A&A...464..429G}; \cite{2013ApJ...772..127T}). 
Recently, the simulations of the radiation 
dominated accretion disks around black holes 
with M-1 closure have been performed by 
\authorcite{2013MNRAS.429.3533S}
(\yearcite{2013MNRAS.429.3533S}, \yearcite{2014MNRAS.439..503S})
and \citet{2014MNRAS.441.3177M}
%
%
%

The hydrodynamical terms for ideal fluid in Eqs. (\ref{cont})-(\ref{gase})
are explicitly solved with using the computational hydrodynamical code,
the Virginia Hydrodynamics One.
The advective transport of radiation energy in Eq. (\ref{rade}) is also 
treated with the explicit method.
In Eqs. (\ref{momphi})-(\ref{rade}),
we solve the gas-radiation interaction terms,
the radiation flux term and the viscous terms 
based on the implicit method.
In order to update the radiation enargy 
and the azimuthal component of the velocity
by solving the radiative flux term and the viscosity term,
we employ the Gauss-Jordan elimination for a matrix inversion.

Our initial and boundary conditions are essentially same as those of O05.
We start the calculations with a hot and rarefied atmosphere,
and matter is continuously injected into the computational 
domain through the outer disk boundary,
$r=r_{\rm out}$ and $0.45\pi\lesssim\theta\lesssim 0.5\pi$
(as we have mentioned above, 
the outer boundary is located at $r=500r_{\rm S}$ in O05
and we employ $r_{\rm out}=5000r_{\rm S}$).
Such injected matter is supposed to 
have a specific angular momentum corresponding to the 
Keplerian angular momentum at $r=r_{\rm rot}$.

In the present study, 
we mainly investigate the case that
the injected mass accretion rate at the outer boundary
(mass input rate, $\dot{M}_{\rm input}$) is $1000 L_{\rm Edd}/c^2$
and $r_{\rm rot}$ is $100r_{\rm S}$ (base line model).
In \S\ref{comparison model},
we discuss the results for 
$\dot{M}_{\rm input}=1000 L_{\rm Edd}/c^2$ 
and $r_{\rm rot}=500r_{\rm S}$ (comparison model 1)
as well as
$\dot{M}_{\rm input}=10000 L_{\rm Edd}/c^2$ 
and $r_{\rm rot}=100r_{\rm S}$ (comparison model 2).

\section{Result}
\label{result}
\subsection{Global inflow-outflow structure}
\begin{figure}[t]
\begin{center}
\includegraphics[clip,width=8cm]{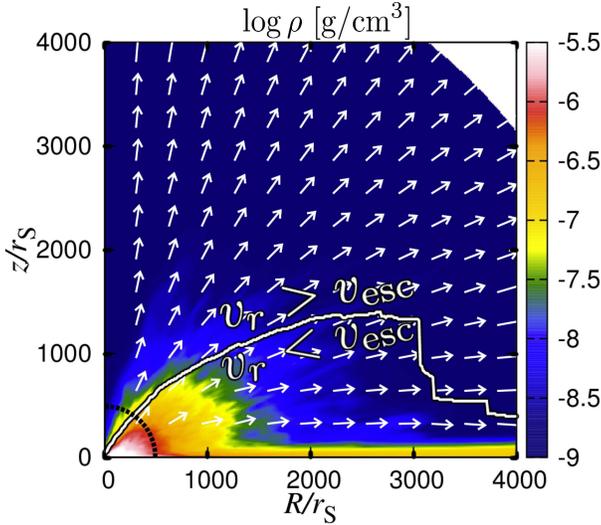}
\caption{
The density distribution (color)
overlaid with the velocity vectors
in quasi-steady state.
They are time-averaged
in the elapsed time between $t=70$ sec and 200 sec.
The white solid line indicates the boundary at which
the radial velocity equals to escape velocity.
Thus, the radial velocity is larger (smaller)
than the escape velocity above (below) the line.
The dotted line indicates the computational domain 
of \cite{Ohsuga2005} (2005).}
\label{fig:dens}
\end{center}
\end{figure}

Although there is no accretion disk initially,
the injected matter accumulates within the computational domain.
Then, the accretion disk as well as the launching outflow appears.
The quasi-steady structure for base line model is shown 
in Fig.\ref{fig:dens},
in which the color contour indicates
the gas density 
and 
arrows mean velocity vectors in $r-z$ plane.
They are time-averaged 
in the elapsed time between $t=70$ sec and 200 sec
(the Keplerian time at $r=r_{\rm rot}$ is about 0.9 sec 
in the present simulations because of $M_{\rm BH}=10M_\odot)$.
The high density region at $r \lesssim 500 r_{\rm S}$
(red and white)
corresponds to super-critical accretion disk (disk region),
which is radiation pressure-dominated, 
and geometrically and optically thick.
The injected matter freely falls
in the yellow region near the equatorial plane
at $r\gtsim 500 r_{\rm S}$ (free-fall region),
since the centrifugal force is less than the gravity.
Except for the disk region and the free-fall region,
we find that the matter moves outwards for the wide angle
(wide-angle outflow region).

The white solid line in this figure indicates
the boundary at which the radial velocity, $v_r$, 
equals to the escape velocity, $v_{\rm esc}$.
Above the line,
the matter is blown away 
with speed of $v_r > v_{\rm esc}$
(high-velocity outflow).
Although we also find outflow motion below the line,
the matter cannot be released from the gravity
(low-velocity outflow).
Fig.\ref{fig:dens} also shows that 
the angular size of the high-velocity outflow 
broadens with an increase of the radius.
The high-velocity outflow appears
only at the angle of $\theta \lesssim 35^\circ$
near the black hole ($r\lesssim 500r_{\rm S}$).
In $\theta \gtsim 35^\circ$,
the matter is gradually accelerated 
and its velocity is over $v_{\rm esc}$ at the outer region.
For example, the radial velocity exceeds escape
velocity at $r\gtsim 1000r_{\rm S}$ for $\theta \sim 45^\circ$,
and we find that $v_r$ achieves $v_{\rm esc}$ 
at $r\sim 3000r_{\rm S}$ for $\sim 70^\circ$.
As a result, the matter is ejected from the outer boundary 
with $v_r>v_{\rm esc}$ for almost all direction
($0^\circ\lesssim\theta\lesssim 85^\circ$)
except for the vicinity of the equatorial plane.

Such a wide-angle high-velocity outflow at the regions of 
$r\gg 500r_{\rm S}$ is for the first time revealed 
by our present study.
O05, in which the computational domain is $500r_{\rm S}$
(dotted line in Fig.\ref{fig:dens}),
showed the high-velocity outflow ($v_r>v_{\rm esc}$)
only near the rotation axis ($\theta \lesssim 35^\circ$).
The radial velocity of $\theta \gtsim 35^\circ$
is less than the escape velocity.
O05 could not reveal the behavior of such low-velocity outflow
at the outer regions because of limited computational domain
(see \S\ref{previous} for detail comparison).
%
%

\begin{figure}[t]
\begin{center}
\includegraphics[clip,width=8cm]{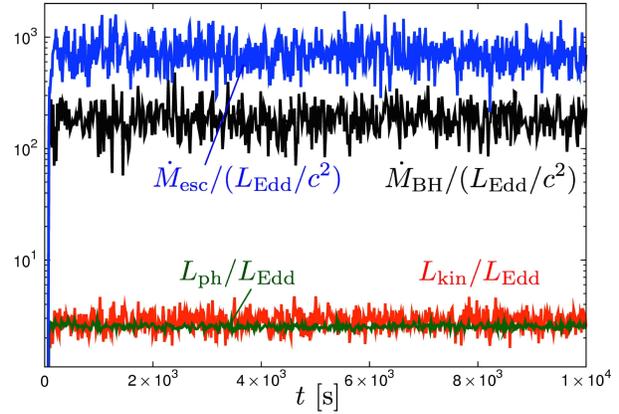}
\caption{
The time evolution of
the normalized photon luminosity ($L_{\rm ph}/L_{\rm Edd}$; green),
kinetic power ($L_{\rm kin}/L_{\rm Edd}$; red),
the mass accretion rate onto the black hole
($\dot{M}_{\rm BH} /(L_{\rm Edd}/c^2)$; black)
and the mass escape rate
($\dot{M}_{\rm esc} /(L_{\rm Edd}/c^2)$; blue).
}
\label{fig:t_evo}
\end{center}
\end{figure}

Our present simulation succeeded in reproducing the
quasi-steady super-critical disk accretion flow.
Fig.\ref{fig:t_evo} shows the time evolution
of the mass accretion rate onto the black hole
(black line),
\begin{equation}
    \dot{M}_{\rm BH}
    =2\pi r_{\rm in}^2
    \int \rho (-v_{\rm r})
    \sin\theta d\theta,
\label{eq:m_bh}
\end{equation}
and the photon luminosity (green line),
\begin{equation}
    {L}_{\rm ph}
    = 2\pi r_{\rm out}^2
    \int F_0^r
    \sin\theta d\theta.
\label{eq:l_ph}
\end{equation}
The mass accretion rate is around 150 times 
larger than the critical accretion rate ($L_{\rm Edd}/c^2$),
and the luminosity is $\sim 2.5 L_{\rm Edd}$
in the quasi-steady state ($t \gtsim 60$ sec).
Such results are roughly consistent with those of O05.
We conclude that the wide-angle outflow does not 
prevent the mass accretion onto the black hole,
and quasi-steady super-critical accretion is feasible.

Fig.\ref{fig:dens} also indicates that
huge amount of matter is blown away by 
the wide-angle high-velocity outflows
and energy is released not only via the radiation
but also via the outflows.
%
The blue line shows the mass escape rate,
\begin{equation}
    \dot{M}_{\rm esc}
    =2\pi r_{\rm out}^2
    \int \rho 
    \left\{
    \begin{array}{ll}
     v_r & {\rm for} \,\,\, v_r\ge v_{\rm esc} \\
     0  & {\rm for} \,\,\, v_r < v_{\rm esc} 
    \end{array}
    \right\}
    \sin\theta d\theta,
\label{eq:m_esc}
\end{equation}
that means the ejected mass per unit time 
through the outer boundary via the high-velocity outflow.
The kinetic power of the high-velocity outflow,
\begin{eqnarray}
  L_{\rm kin}&=&2\pi r_{\rm out}^2 \nonumber\\
   &\times &\int \left(\frac{1}{2}\rho v^2\right)
       \left\{
    \begin{array}{ll}
     v_r & {\rm for} \,\,\, v_r\ge v_{\rm esc} \\
     0  & {\rm for} \,\,\, v_r < v_{\rm esc} 
    \end{array}
    \right\}
  \sin\theta d\theta,
\end{eqnarray}
are also plotted in Fig.\ref{fig:t_evo} (red).
It is found that both the mass escape rate and 
the kinetic power is quasi-steady.
We find $\dot{M}_{\rm esc} \sim 700 L_{\rm Edd}/c^2$,
which corresponding to $70\%$ of the mass input rate
and largely exceeds the mass accretion rate onto the black hole, 
$\dot{M}_{\rm BH}\sim 150 L_{\rm Edd}/c^2$.
We confirmed that gas is ejected 
from the super-critical accretion flows
with the rate of $\dot{M}_{\rm esc}>\dot{M}_{\rm BH}$.
We also find that the kinetic power 
of the high-velocity outflow
is over the Eddington luminosity
and comparable to the photon luminosity,
$L_{\rm kin}\sim L_{\rm ph}$.

O05 indicated $\dot{M}_{\rm esc}\sim 100L_{\rm Edd}/c^2$.
However this value was evaluated at 
their outer boundary, $r_{\rm out}=500r_{\rm S}$.
Since the matter is accelerated at the regions of $r>500r_{\rm S}$,
the mass escape rate 
is much larger in the present simulation than that in O05.

\subsection{Radiation force}
\begin{figure}[t]
\begin{center}
\includegraphics[clip,width=8cm]{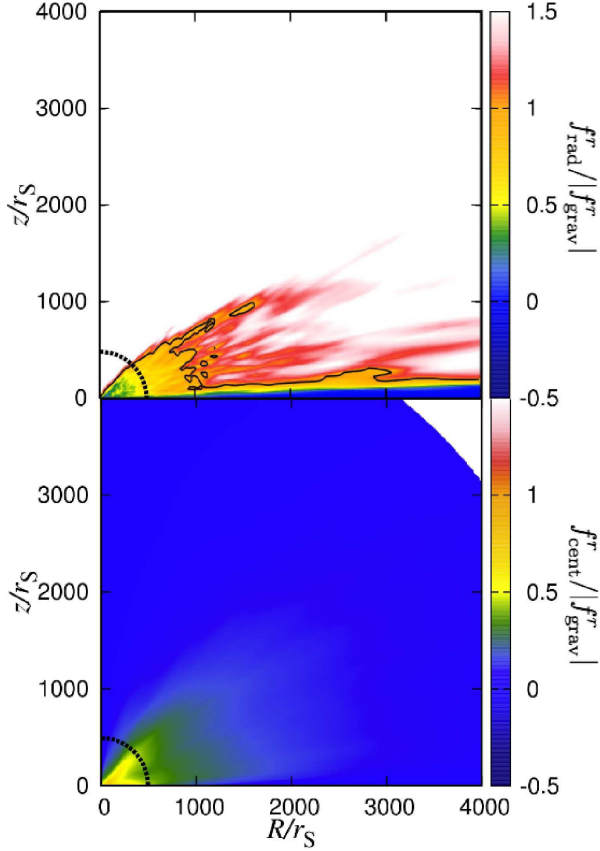}
\caption{Upper panel:
the ratio of the radial component of the radiation force and the gravity.
The solid line indicates a contour of unity
(The outward radiation force equals the gravity). 
Lower panel: the radial component of the centrifugal force
divided by the gravity. Both are time-averaged
in the elapsed time between $t=70$ sec and 200 sec.
}
\label{fig:frad_fcent}
\end{center}
\end{figure}

Here we show that the
wide-angle outflow is driven via
the radiation force.
It is clearly understood by Fig.\ref{fig:frad_fcent}.
The upper panel of this figure indicates
the ratio of the radial component of the radiation force 
($f^r_{\rm rad}=\chi F^r_0/c$) and the gravity
($|f^r_{\rm grav}|=GM/[r-r_{\rm S}]^2$).
In the lower panel, 
we plot the radial component of the centrifugal force
($f^r_{\rm cent}=v^2_\varphi/r$) divided by the gravity.
Here, they are also time-averaged over 130 sec ($70-200$ sec).
We find in the upper panel that 
the radiation force is dominant over the gravity
in the wide-angle outflow region (see white and red).
On the other hand, 
the centrifugal force is less effective at this region
(blue in the lower panel).
Since the gas pressure force is also much smaller than
the radiation force,
we conclude that the wide-angle outflow
is mainly driven by the radiation force.

Around the rotation axis,
since the radiation force exceeds the gravity near the black hole,
the high-velocity outflow emerges from the inner region.
In contrast, in the direction of $\theta \gtsim 45^{\circ}$,
the radiation force exceeds the gravity only at the outer region,
$r\gtsim 1000 r_{\rm S}$.
Thus, the angle of $v_r>v_{\rm esc}$ gradually broadens
with an increase of the radius,
producing the wide-angle high-velocity outflow at $r \gtsim 4000r_{\rm S}$
(see Fig.\ref{fig:dens}).

Here, we note that 
the radiation force is less effective
against freely falling matter near the 
equatorial plane,
since the radiation does not penetrate
the very optically thick material.
Therefore, the radiation pressure does not prevent 
the free-falling motion along the equatorial plane.
Indeed, we can see that the 
radiation force is very small at 
$\theta\gtsim 85^\circ$ and $r\gtsim 1000r_{\rm S}$
in the upper panel of Fig.\ref{fig:frad_fcent} (blue).
In the disk region ($r\lesssim 500r_{\rm S}$), 
both of the centrifugal force and the radiation force is
less than the gravity
(shown in yellow and green in the upper and lower panels, respectively).
However, sum of them are roughly balances with the gravity.
Thus, the matter slowly accretes onto the black hole,
since the angular momentum is transported via the viscosity.
Such a feature has been shown by \cite{Ohsuga-Mineshige2007}(2007).

\subsection{Structure of wide-angle outflow}
\begin{figure}[t]
\begin{center}
\includegraphics[clip,width=8cm]{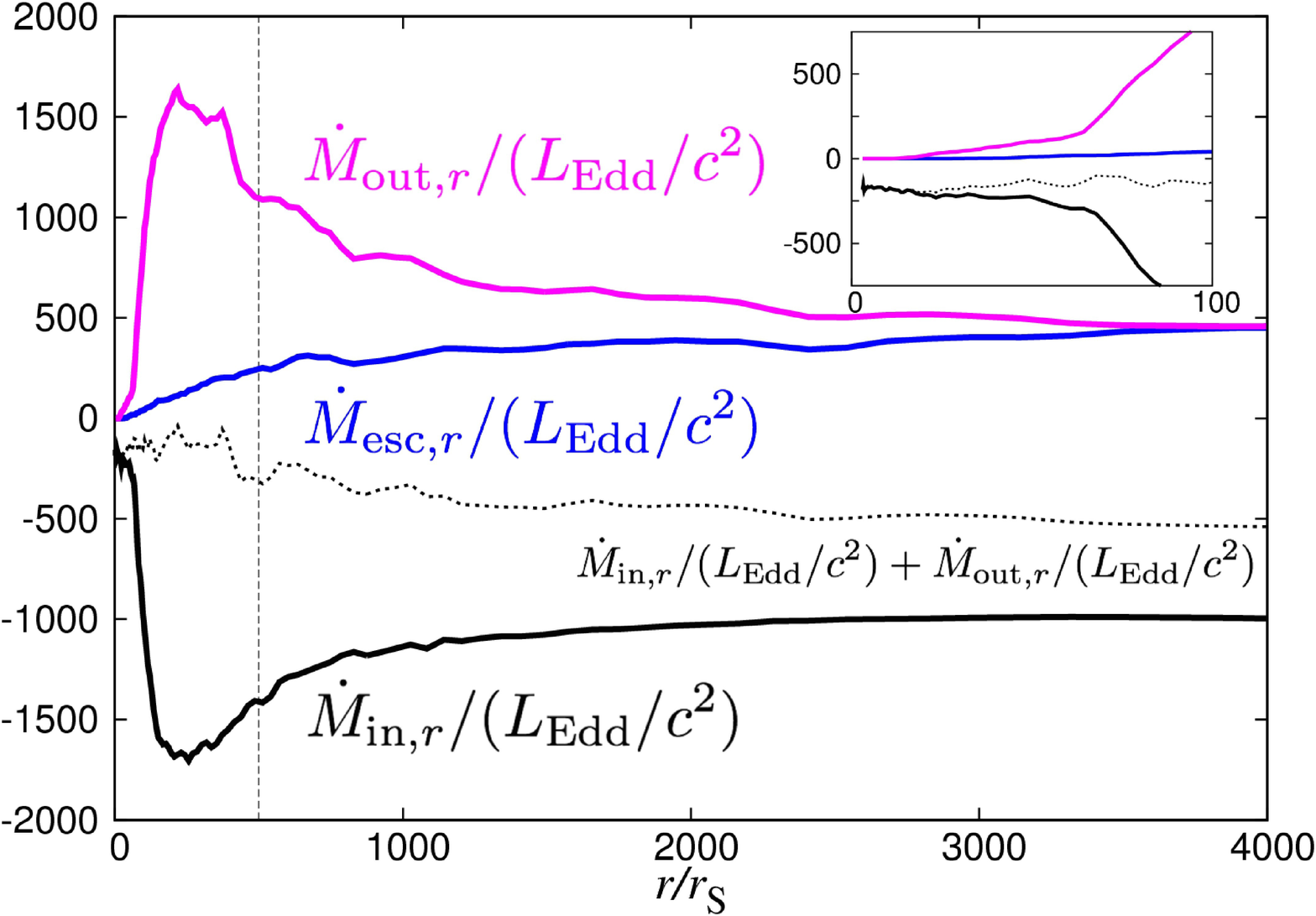}
\caption{
The $r$-dependent the mass escape rate (blue),
the mass outflow rate (magenta),
the mass inflow rate (black),
and the sum of mass inflow rate and mass outflow rate (dotted).
All values are time-averaged over 130 sec ($70-200$ sec).
The profiles of them naer the black hole are shown in the small panel.
}
\label{fig:r-dependent}
\end{center}
\end{figure}

Next, we quantitatively discuss the outflow motion, 
as well as accretion motion.
We plot in Fig.\ref{fig:r-dependent} 
the $r$-dependent mass escape rate (blue),
\begin{equation}
    \dot{M}_{{\rm esc},r}
    =
    \int 2\pi r^2 \rho 
    \left\{
    \begin{array}{ll}
     v_r & {\rm for} \,\,\, v_r\ge v_{\rm esc} \\
     0  & {\rm for} \,\,\, v_r < v_{\rm esc} 
    \end{array}
    \right\}
    \sin\theta d\theta,
\label{eq:m_escr}
\end{equation}
mass outflow rate (magenta),
\begin{equation}
    \dot{M}_{{\rm out},r}= 
    \int2\pi r^2\rho \max[v_r,0] \sin\theta d\theta,
\label{eq:m_out}
\end{equation}
and mass inflow rate (black),
\begin{equation}
    \dot{M}_{{\rm in},r}=
    \int2\pi r^2\rho \min[v_r,0] \sin\theta d\theta.
\label{eq:m_in}
\end{equation}
Note that the mass accretion rate onto the black hole, $\dot{M}_{\rm BH}$,
and the mass escape rate, $\dot{M}_{\rm esc}$,
correspond to $\dot{M}_{{\rm in}, r}$ and $\dot{M}_{{\rm esc},r}$
at $r=r_{\rm in}$ and $r=r_{\rm out}$, respectively
(see Eqs. [\ref{eq:m_bh}] and [\ref{eq:m_esc}]).
The dotted line indicates the sum of the $r$-dependent
mass inflow rate and mass outflow rate, 
$\dot{M}_{{\rm in}, r}+\dot{M}_{{\rm out}, r}$.
All values are time-averaged over 130 sec ($t=70-200$ sec).

As shown in Fig.\ref{fig:r-dependent},
the mass escape rate monotonically 
increases with an increase of the radius.
This is because that 
the radiation force continues to accelerate the matter,
increasing the mass of the outflow with $v_r>v_{\rm esc}$.
Although the mass escape rate is smaller than
the mass outflow rate,
$\dot{M}_{{\rm esc}, r}<\dot{M}_{{\rm out}, r}$, within $r\sim 3500r_{\rm S}$,
we find $\dot{M}_{{\rm esc}, r} \sim \dot{M}_{{\rm out}, r}$ 
at $r\gtrsim 3500r_{\rm S}$.
This implies that the velocity of almost all outflowing matter 
exceeds the escape velocity,
and all the matter ejected from the computational domain
is blown away far in the distance.
Here we note that 
the mass escape rate as well as outflow rate
at the outer boundary
goes up $\sim 700 L_{\rm Edd}/c^2$ when $t\gtrsim 500$ sec
in spite of $\dot{M}_{{\rm out}, r} \sim \dot{M}_{{\rm esc}, r} 
\sim 500 L_{\rm Edd}/c^2$ in Fig.\ref{fig:r-dependent}.

In the regions of $r\lesssim 700r_{\rm S}$,
we find $\dot{M}_{{\rm out}, r}>10^3 L_{\rm Edd}/c^2$ 
and $\dot{M}_{{\rm in}, r}<-10^3 L_{\rm Edd}/c^2$.
This is caused by the convective motion or circulation
of the high-density gas in the disk region.
The sum of the mass inflow rate and outflow rate
is somewhat negative, implying that 
the disk matter gradually accretes towards the black hole.
The flat profile of $\dot{M}_{{\rm in}, r} (\sim -10^3 L_{\rm Edd}/c^2)$
at $r\gtsim 1500r_{\rm S}$
indicates that the injected matter freely falls 
along the equatorial plane.

Here, we note that sum of $-\dot{M}_{{\rm in}, r} (r=r_{\rm in})$
and $\dot{M}_{{\rm out}, r} (r=r_{\rm out})$
is about $70\%$ of $-\dot{M}_{{\rm in}, r} (r=r_{\rm out})$.
It means that the $70\%$ of the injected matter 
goes out from the computational domain 
and total mass in the domain 
continuously increases with time. 
The matter seems to guradually 
accumurate at around $r\sim 500-1500 r_{\rm S}$,
since the gradient of the dotted line, 
$\dot{M}_{{\rm in}, r}+\dot{M}_{{\rm out}, r}$, is negative.
In contrast, the flat profile 
of the dotted line near the black hole and outer region 
($r\gtsim 1500r_{\rm S}$) implies that 
the inflow/outflow equiliblium is achieved.
Although we find that 
$\dot{M}_{{\rm out}, r} (r=r_{\rm out})
-\dot{M}_{{\rm in}, r} (r=r_{\rm in})$
becomes $\sim 90\%$ of mass input rate
at $t\gtrsim 500$ sec
(see Fig.\ref{fig:t_evo}),
we need long-term calculations 
in order to examine a conclusive steady structure,
$\dot{M}_{{\rm out}, r} (r=r_{\rm out})
-\dot{M}_{{\rm in}, r} (r=r_{\rm in})
=\dot{M}_{{\rm in}, r} (r=r_{\rm out})$.

\begin{figure}[t]
\begin{center}
\includegraphics[clip,width=8cm]{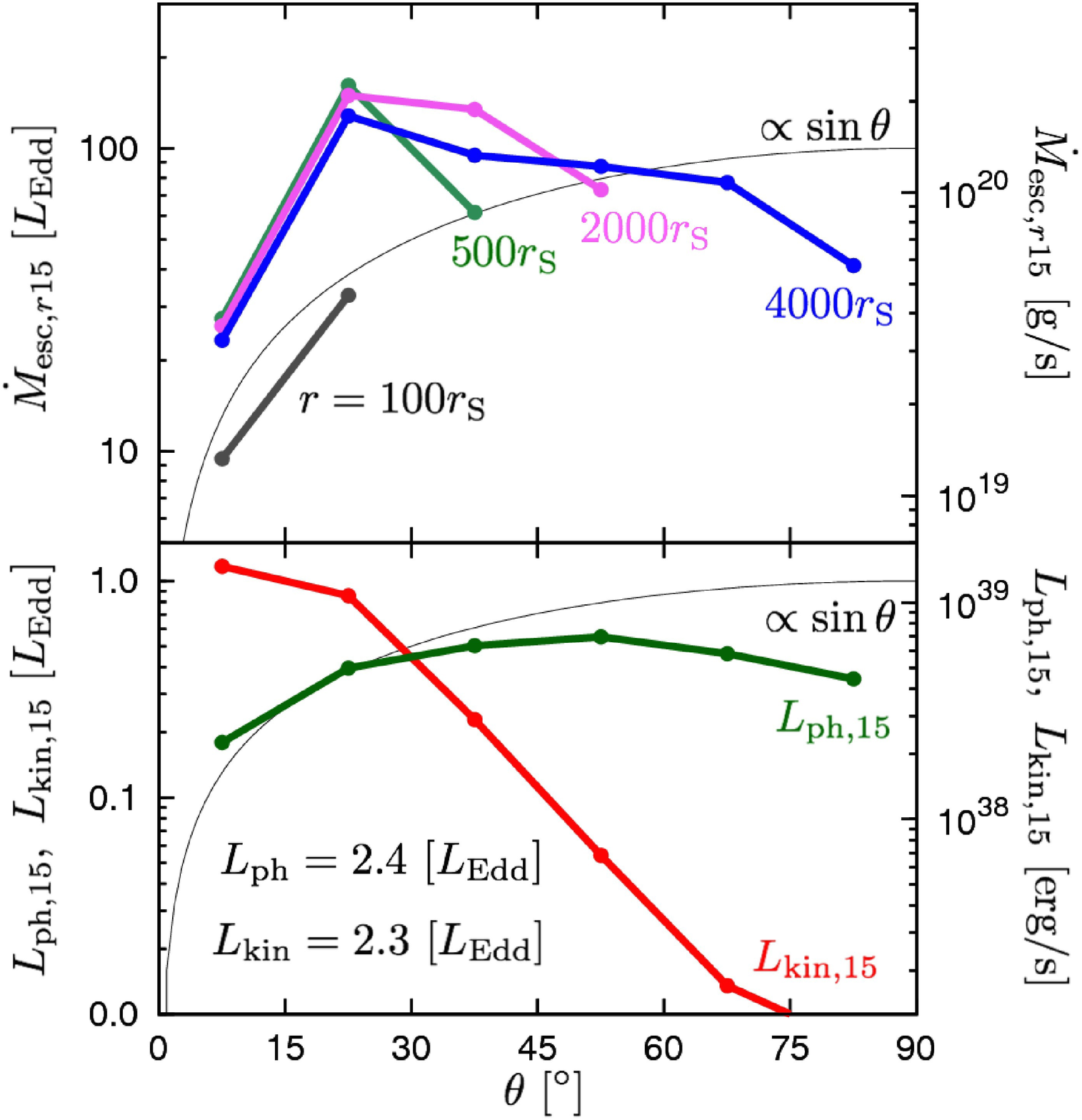}
\caption{
Upper panel:
mass transported outwards per unit time
via the high-velocity outflow
as a function of angle
for $100r_{\rm S}$ (gray), $500r_{\rm S}$ (light green),
$2000r_{\rm S}$ (light magenta), and $4000r_{\rm S}$ (blue).
Lower panel: kinetic energy transported outwards
via the high-velocity outflow
per unit time
as a function of the angle (red).
The radiation flux integrated
with respect to each $15^\circ$
is also plotted (green).
All values are time-averaged over 130sec ($70-200$ sec).
}
\label{fig:mescr_power}
\end{center}
\end{figure}

The upper panel of Fig.\ref{fig:mescr_power} shows
the $r$-dependent mass escape rate 
split into an angular range of $15^\circ$,
\begin{eqnarray}
 \dot{M}_{{\rm esc}, r15}
  = 
  2\pi &&
  \int^{\theta+7.5^\circ}_{\theta-7.5^\circ}
  r^2\rho \nonumber \\
  &&\times
   \left\{
   \begin{array}{ll}
    v_r & {\rm for} \,\,\, v_r\ge v_{\rm esc} \\
    0  & {\rm for} \,\,\, v_r < v_{\rm esc} 
   \end{array}
  \right\}
  \sin\theta d\theta,
\end{eqnarray}
for $r=100r_{\rm S}$ (gray), $500r_{\rm S}$ (light green),
$2000r_{\rm S}$ (light magenta), and $4000r_{\rm S}$ (blue).
Here, this rate is time-averaged
in the elapsed time between $t=70$ sec and 200 sec.

We find in this panel that
the velocity of outflow is larger than the escape velocity 
near the rotation axis in the vicinity of the black hole.
The angular size of high-velocity outflow
tends to broaden with an increase of the radius.
In particular, 
although the high-velocity outflow appears 
only in the direction of $\theta\ltsim 35^\circ$
at $r=100r_{\rm S}$,
the angular size of the high-velocity outflow extends 
up to $\sim 20^\circ$ and $\sim 50^\circ$
at $r=500r_{\rm S}$ and $2000r_{\rm S}$.
Eventually, the matter is blown away with speed of $>v_{\rm esc}$
in the wide angle, $0-85^\circ$ (see $r=4000r_{\rm S}$).
We find that $\dot{M}_{{\rm esc},r15}$ at $r=4000r_{\rm S}$
is not so sensitive to the angle. 
It is slightly small near the rotation axis and equatorial plane
and slightly enhanced around $20^\circ$.
Note that mass flux via the high-velocity is mildly collimated 
(compare with sine curve [solid line]).
We conclude that the mass is blown away 
with speed of $>v_{\rm esc}$ towards all direction
somewhat focusing around $20^\circ$.

In the lower panel of Fig.\ref{fig:mescr_power},
we plot kinetic power and photon luminosity
split into an angular range of $15^\circ$,
\begin{eqnarray}
 L_{{\rm kin}, 15}
  = 2\pi r_{\rm out}^2 &&
  \int^{\theta+7.5^\circ}_{\theta-7.5^\circ} 
  \left(
  \frac{1}{2}\rho v^2
  \right) \nonumber\\
  &&\times \left\{
   \begin{array}{ll}
    v_r & {\rm for} \,\,\, v_r\ge v_{\rm esc} \\
    0  & {\rm for} \,\,\, v_r < v_{\rm esc} 
   \end{array}
   \right\}
   \sin\theta d\theta,
  \label{eq:L_kin}
\end{eqnarray}
and
\begin{equation}
 L_{{\rm ph}, 15}
  = 2\pi r_{\rm out}^2
  \int^{\theta+7.5^\circ}_{\theta-7.5^\circ}
  F^r_0 \sin\theta d\theta,
\label{eq:L_obs}
\end{equation}
time-averaged in the elapsed time between $t=70$ sec and 200 sec.
In addition to the mass, 
it is found that the kinetic energy is released in all direction
except for the vicinity of the equatorial plane, $\theta\gtsim 85^\circ$.
However, in contrast that $\dot{M}_{{\rm esc}, r15}$ 
for $r=4000r_{\rm S}$ is small at $\theta \lesssim 15^\circ$,
$L_{{\rm kin}, 15}$ is enhanced around the rotation axis.
We find that $L_{{\rm kin}, 15}$ is 100 times larger 
for $\theta \sim 10^\circ$ than for $\theta \sim 80^\circ$.
We also find that 
$L_{{\rm ph}, 15}$ is not so sensitive to the angle
(Note that the radial component of the radiative flux,
$F_0^r$, is mildly collimated [compare with sine curve]).
As we have already mentioned,
the resulting total kinetic power is comparable to 
the photon luminosity, $L_{\rm kin} \sim L_{\rm ph}$.
Since we set the black hole mass to be $10M_\odot$,
we have $L_{\rm kin} \sim 3\times 10^{39}\ [{\rm erg/s}]$.

\section{Discussion}
\label{discussion}
\subsection{Comparing with \cite{Ohsuga2005}(2005)}
\label{previous}
Here, we compare our results with O05.
In O05,
although the huge amount of the gas is ejected from 
the their computational domain, $r=500r_{\rm S}$,
the velocity of the outflowing matter 
does not exceed the escape velocity
except for the high-velocity outflow near the rotation axis.
At the outer boundary, mass outflow rate is about 
several $100 L_{\rm Edd}/c^2$ 
but $\dot{M}_{\rm esc} \sim 100 L_{\rm Edd}/c^2$.
Hence, it is not clear that 
whether such a low-velocity outflow come back to
the neighborhood of the black hole 
or is blown away due to the radiation force.
In the present study, 
the occurrence of the wide-angle high-velocity outflow 
is for the first time revealed.
The low-velocity outflow is accelerated at the outer region
and its velocity exceeds the escape velocity at
$r\gtsim 4000r_{\rm S}$.

\begin{figure}[t]
\begin{center}
\includegraphics[clip,width=8cm]{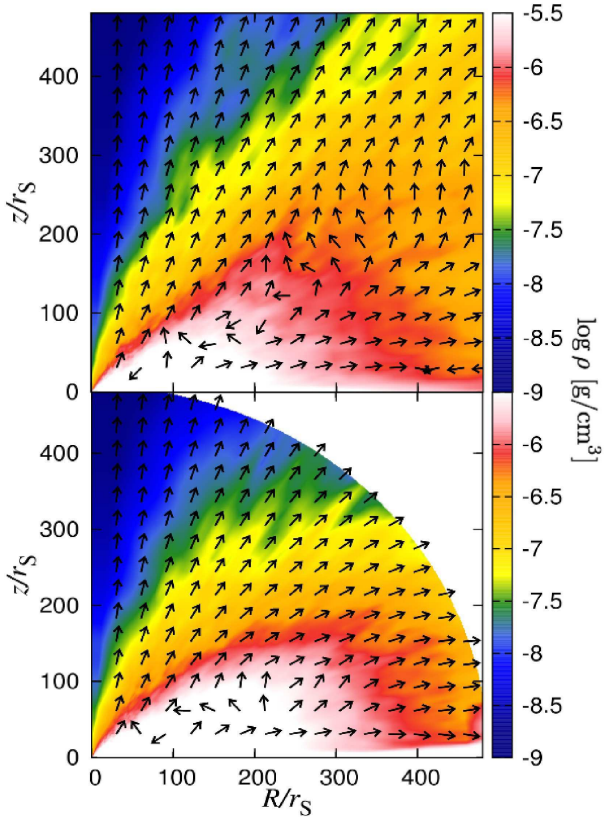}
\caption{
Time averaged density distribution
overlaid with the time-averaged velocity vectors
of the present simulation (upper panel)
and O05 (lower panel).
}
\label{fig:dens_compare}
\end{center}
\end{figure}

Our resulting flow structure is distinct from 
that of O05 at around $r=300-500 r_{\rm S}$.
Upper and lower panels in Fig.\ref{fig:dens_compare} 
display the time-averaged (40 sec) density distribution
overlaid with the velocity vectors
of the present work and O05.
There is no remarkable difference
in the vicinity of the black hole
($r\lesssim 100r_{\rm S}$).
However, high-density region (white and yellow) is wider
in the present study than in O05, 
(see the region of $r\sim 300-500r_{\rm S}$).
This is caused by that
the outer boundary is located at $r=500r_{\rm S}$ in O05.
In our simulations,
although the time-averaged velocity is outward as shown
in the top panel, circular motion occurs around
$r\sim 300-700r_{\rm S}$.
In contrast,
since the matter that goes out through the boundary
can not come back,
the small domain in O05
tends to decrease the density.

\subsection{Wide-angle outflows of comparison models}
\label{comparison model}
\begin{figure}[t]
\begin{center}
\includegraphics[clip,width=8cm]{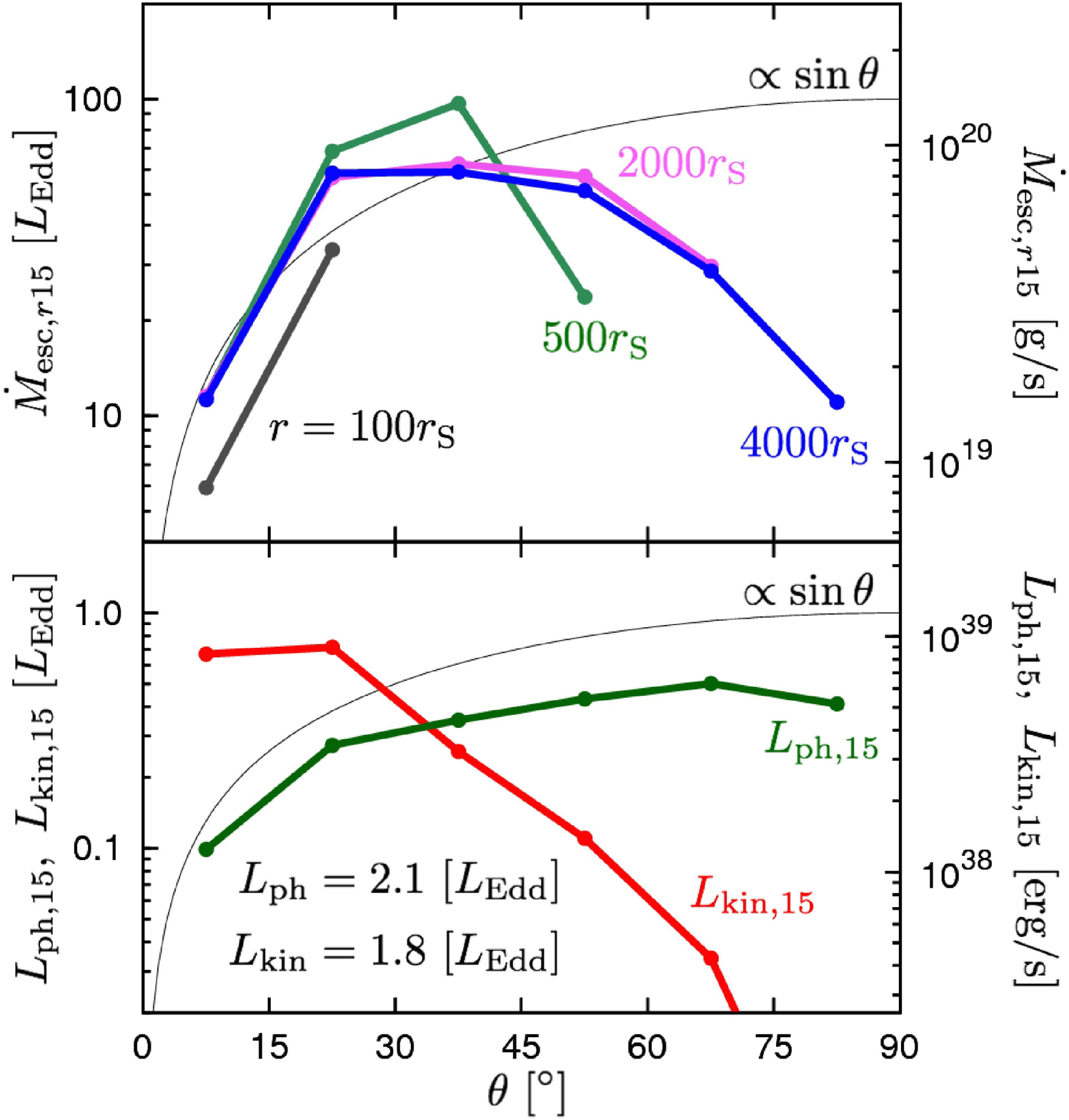}
\caption{Same as Fig.\ref{fig:mescr_power}
but for comparison model 1,
in which we employ $r_{\rm rot}=500r_{\rm S}$.
}
\label{fig:model1}
\end{center}
\end{figure}

Next, we show the results of comparison models 1 and 2.
In comparison model 1, we employ $r_{\rm rot}=500r_{\rm S}$.
Then, the centrifugal force tends to prevent the free-falling motion
around $500r_{\rm S}$.
Although the disk region expands in the horizontal direction,
the super-critical flow is realized.
Resulting photon luminosity is $\sim 2 L_{\rm Edd}$
and the mass accretion rate onto the black hole is 
$\sim 100 L_{\rm Edd}/c^2$.
We plot $\dot{M}_{{\rm esc}, r15}$, 
$L_{{\rm kin}, 15}$, and $L_{{\rm ph}, 15}$ 
in Fig.\ref{fig:model1}.
In the upper panel of this figure, 
we find that the angular size of the high-velocity outflow
increases with an increase of the radius,
and wide-angle high-velocity outflow ($\theta \lesssim 85^\circ$)
appears at $r\gtsim 4000r_{\rm S}$.
We also find that $L_{{\rm kin}, 15}$ is 
mildly collimated in contrast with $L_{{\rm ph}, 15}$.
Such tendency is basically similar with the base line model.
The kinetic power and the photon luminosity of this model 
are $1.8L_{\rm Edd}$ and $2.1L_{\rm Edd}$, respectively.
These values are also similar to those in the base line model.

\begin{figure}[t]
\begin{center}
\includegraphics[clip,width=8cm]{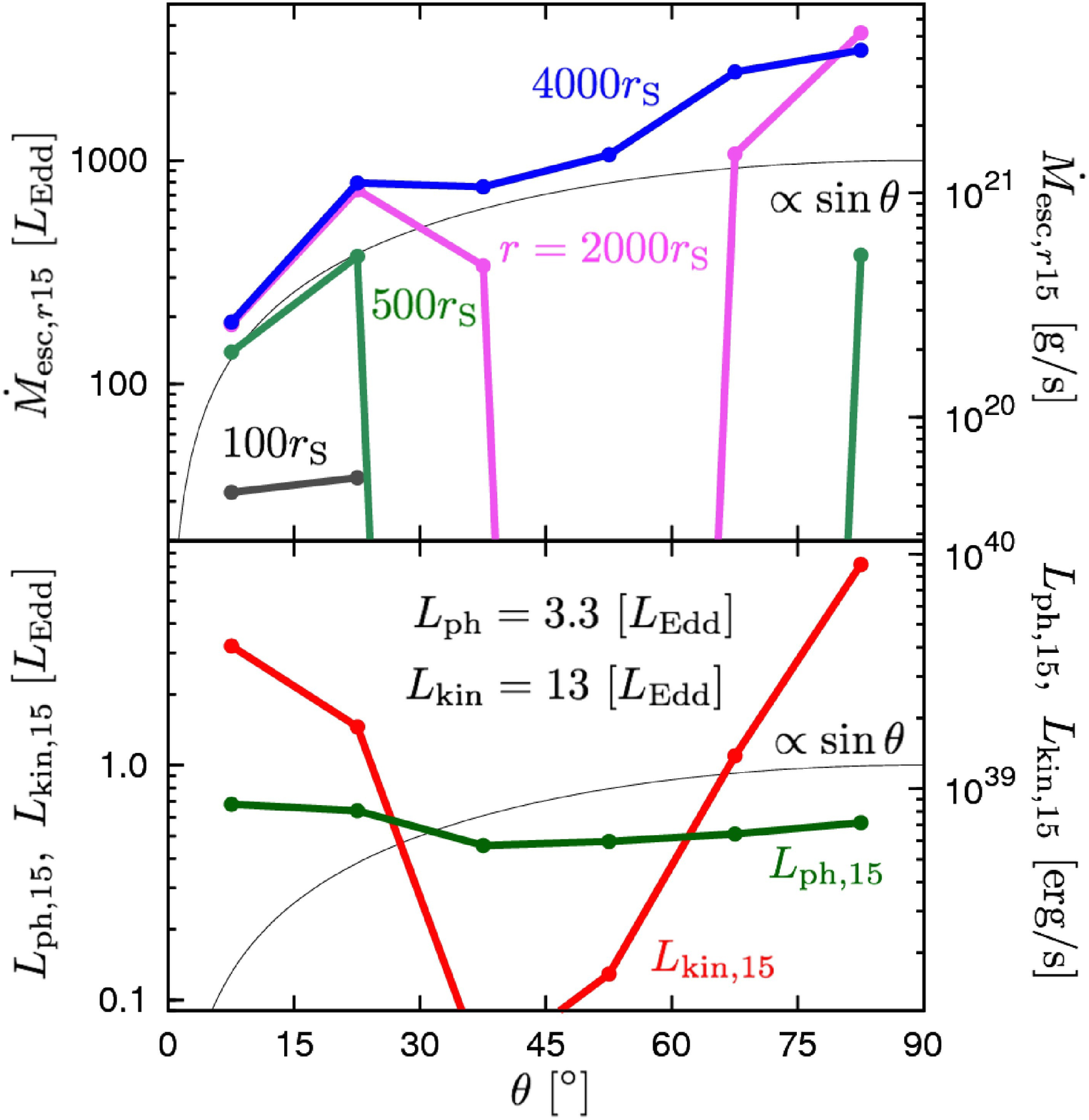}
\caption{
Same as Fig.\ref{fig:mescr_power}
but for comparison model 2,
in which we suppose the mass input rate
to be $1000 L_{\rm Edd}/c^2$.
}
\label{fig:model2}
\end{center}
\end{figure}

In the comparison model 2, we set the mass input rate to be
$10000 L_{\rm Edd}/c^2$ and $r_{\rm rot}$ is kept $100r_{\rm S}$.
In this model, global structure of the outflow drastically changes.
Huge amount of the matter is injected within $r=100r_{\rm S}$
and mass accretion rate onto the black hole becomes 
very large, $\dot{M}_{\rm BH} \sim 700 L_{\rm Edd} /c^2$.
Then, the enhanced radiation force
in cooperation with the centrifugal force
blows away the disk matter in the horizontal direction
(horizontal outflow).
In the top panel in Fig.\ref{fig:model2}, 
we find $\dot{M}_{{\rm esc}, r15}$
is enhanced in the direction of $\theta\gtsim 70^\circ$.
This horizontal outflow raises $L_{{\rm kin}, r15}$ 
for $\theta\gtsim 70^\circ$ (see bottom panel
in Fig.\ref{fig:model2}),
leading to the large kinetic power of 
$L_{\rm kin} \sim 13 L_{\rm Edd}$, 
which is larger than the photon luminosity.
However, such horizontal outflow might disappear 
if we employ larger $r_{\rm rot}$,
since, then, the disk region expands 
in the horizontal direction and 
the horizontal radiation force 
is reduced due to the large optical thickness 
of the disk matter.
In order to make clear the point,
we should perform the simulations 
with $r_{\rm rot} \gg 1000r_{\rm S}$.
Such simulations require much longer computation times
because of long viscosity timescale.

\subsection{Bubbles around ultra-luminous X-ray sources}
\label{bubbles}
By our global RHD simulations of super-critical flows around 
stellar mass black holes,
we revealed the wide-angle, powerful outflows launched from
the accretion disks via the radiation force.
Such outflows would evolve into the bubbles 
as observed around some ULXs.

Some bubbles around ULXs 
are thought to be excited by the shock/outflow 
(\cite{Pakull-Mirioni2002}2002; \cite{Pakull-Mirioni2003}2003;
\cite{Grise-Pakull-Motch2006}2006; \cite{Abolmasov2007}2007;
\cite{Cseh2012}2012).
Recently, \cite{Cseh2012}(2012) estimated the kinetic power 
required for the shock-excited bubbles associated with a ULX, IC 342 X-1,
to be $\sim 3 \times 10^{39}\ {\rm erg\ s^{-1}}$
using the bubble theory (\cite{Weaver1977}1977).
This is consistent with the kinetic power seen in our model.
In addition, the observed X-ray luminosity of the central source
is the same order, $\sim 5 \times 10^{39}\ {\rm erg\ s^{-1}}$,
which is also consistent with our model, where 
the radiation luminosity and kinetic power are comparable.
Our results nicely fit the observed properties of bubbles around ULXs.
And therefore, our model supports a hypothesis that 
ULXs are powered by the super-critical accretion 
onto stellar mass black holes.

Another hypothesis of ULXs is the intermediate mass black holes (IMBHs) 
surrounded by sub-critical accretion disks.
Then, large X-ray luminosity can be explained 
since the Eddington luminosity is 
about $10^{41} {\rm erg\ s^{-1}}$ for $M_{\rm BH}=1000M_\odot$.
However, driving mechanism of the powerful outflows
from the sub-critical disks are not understood yet.
Also, the mechanism of energy conversion 
from the X-ray to the outflow is not resolved yet.

Interestingly, \cite{Cseh2014}(2014) found a
jet-like structure around a ULX Holmberg II X-1.
Since the nebula itself is not always excited by the shock,
the direct comparison may be difficult.
Nevertheless, the observed bipolar structure is reminiscent of 
what is seen in our model (Fig. \ref{fig:mescr_power}),
where the kinetic power is higher around the polar direction.
The required kinetic power for the observed jet component 
($\sim 2 \times 10^{39}\ {\rm erg\ s^{-1}}$)
is also similar to that obtained in our model.
It is noted that such a structure is found not only around ULXs,
but also around a microquasar, NGC 7793 S26,
for which the super-critical accretion has been suggested
(\cite{Pakull-Soria-Motch2010}2010; \cite{Soria2010}2010).

Some of ULXs show the time variation in X-ray band.
\cite{Middleton2013}(2013) suggested that
such variation is induced by the clumpy structure
in the disk wind.
In their model, ULXs exhibit the soft spectra
and time variation for off-axis observers.
In contrast, face-on observes detect
the hard spectra and weak time variation
(see also Sutton et al. 2013a; 2013b).
Although the present work focus on the time-averaged structure,
our simulations show the messy structure in the wide-angle outflow.
This is also consistent with the observations of ULXs.
However, spatial resolution in the present study
is not enough to resolve the 
clumpy structure in detail.
\cite{Takeuchi2013}(2013)
demonstrated clumpy disk wind launched from the super-critical 
disks by high-resolution simulations
 (see also \cite{Takeuchi2014}2014).
Thus, we stress again that 
our model supports the model of the super-critical accretion
onto the stellar mass black holes in ULXs.

\section{Conclusions}
\label{conc}
By performing two-dimensional RHD simulations
with large computational box, $r=5000r_{\rm S}$,
we investigate the wide-angle, powerful outflow 
launched from the super-critical accretion disks.
The strong radiation force drives 
the disk wind out in all directions
($\theta \sim 0-85^\circ$)
except for the vicinity of the equatorial plane,
in which the matter accretes inwards.
The angular size of the high-velocity outflow,
of which the outflowing velocity is larger than
the escape velocity,
broadens with an increase of the distance
from the black hole.
The outflow around the rotation axis is 
effectively accelerated and exceeds the 
escape velocity near the black hole.
In contrast, at the larger polar angle,
the matter continues to be gradually accelerated 
and attains the escape velocity at a distance
of a few thousand Schwarzschild radius.
Since the high-velocity outflow
does not prevent the mass accretion
along the equatorial plane,
quasi-steady super-critical accretion onto the black hole
is realized.

Due to the wide-angle high-velocity outflow,
the mass and kinetic energy is ejected
in all directions.
However, the kinetic power is larger
in the polar direction 
than in the horizontal direction.
The kinetic power 
is comparable to the photon luminosity,
and is a few times larger than the Eddington luminosity,
which corresponds 
to $\sim 10^{39}-10^{40}\ {\rm erg\ s^{-1}}$
for the stellar mass black holes.
Our results nicely fit the recent observations of 
shock-excited bubbles around ULXs.
Thus, our model supports a hypothesis that 
ULXs are powered by the super-critical accretion 
onto stellar mass black holes.

\bigskip
\bigskip
We would like to thank the anonymous reviewer for many helpful comments.
Numerical computations were carried out
on XT4 and XC30 system at the Center for Computational
Astrophysics (CfCA) of National Astronomical Observatory
of Japan. 
This work is supported by Grant-in-aids from the Ministry of Education,
Culture, Sports, Science, and Technology (MEXT) of Japan,
No. 24740127 (KO), No. 24740117, 25103515 (MT).


\end{document}